\documentclass[RNAAS]{aastex631}

\usepackage[utf8]{inputenc}
\usepackage{xcolor}

\usepackage{rotating}
\begin{document}

\received{20211229}
\accepted{20220103}

\shortauthors{Hillenbrand \& Rodriguez}

\correspondingauthor{Lynne A. Hillenbrand}
\author{Lynne A. Hillenbrand}
\email{lah@astro.caltech.edu}
\affiliation{California Institute of Technology, Department of Astronomy \\
1200 East California Blvd\\
Pasadena, CA, 91125, USA}

\author[0000-0003-4189-9668]{Antonio C. Rodriguez}

\affiliation{California Institute of Technology, Department of Astronomy \\
1200 East California Blvd\\
Pasadena, CA, 91125, USA}

\title{Expected FU Ori Outburst Amplitudes from the Optical to the Mid-Infrared}

\begin{abstract}
Disks around young stellar objects (YSOs) consist of material that
thermally emits the energy provided by a combination of 
passive heating from the central star, and active, viscous heating due to mass accretion.
FU Ori stars are YSOs with substantially enhanced accretion rates in their inner disk regions.
As a disk transitions from standard low-state, to FU Ori-like high-state accretion,
the outburst manifests through photometric brightening over a broad range of wavelengths. 
We present results for the expected amplitudes of the brightening 
between $\sim$4000 \AA\ and 8 $\mu$m -- the wavelength range where FU Ori type outburst events
are most commonly detected. Our model consists of an optically thick passive $+$ active 
steady-state accretion disk with low and high accretion states. 
\end{abstract}

\section{Introduction}
FU Ori stars are young stellar objects (YSOs) that are currently in outburst, 
driven by enhanced accretion from a protoplanetary disk onto a central young star \citep{hartmann1996}.    
While photometric variability is a general empirical feature of YSOs,
with much of that variability driven by accretion-related phenomena occurring near the star-disk magnetospheric region, 
the accretion burst and large-amplitude outbursts of some YSOs put them in a distinct category.
Currently there are several poorly defined sub-categories of burst and outburst behavior.  
Discrete brightening events can last from days to weeks (the bursts), to months and decades, 
with generally larger amplitudes (outbursts) for the longer duration events.

The most extreme outbursts, known as FU Ori objects, 
are commonly interpreted as the result of disk instabilities driving enhanced accretion and leading to 
crushing of the magnetospheric region  that otherwise channels the accretion.  
The innermost disk geometry likely transitions during an outburst into a more standard inner accretion disk,
with a classical boundary layer between the disk and the star, as in CVs and other compact object accretors
(though no such boundary layer has been observed in any FU Ori star). 

Several of the most famous FU Ori's are optically visible (FU Ori, V1057 Cyg, V1515 Cyg) 
and the observational properties of these early prototypes have long served, over the past five decades, 
to guide the definition  and interpretation of the FU Ori class. 
A salient characteristic is detection of a large-amplitude (4-6 mag) optical outburst, with a rise time of months to years.
The FU Ori class later expanded to also include sources in which 
a substantial near-infrared (rather than optical) brightening was detected.
Additional candidates are those in which no outburst was documented
through observation, but the source presents a spectrum that is ``FU Ori-like".
At least half to two-thirds of the presently claimed FU Ori population is embedded, with $A_V > 5$ mag \citep{connelley2018}.

\section{Identification of FU Ori Outbursts in Photometric Surveys}

In recent years, optical time domain surveys (e.g. PTF/ZTF, ASAS-SN, Gaia, ATLAS)
have continued to detect large-amplitude, long timescale photometric events that can be spectroscopically followed up
and confirmed as analogs to the traditionally defined FU Ori sources.   Examples over the past decade include 
V2493 Cyg (HBC 722), V960 Mon, Gaia 17bpi, and Gaia 18dvy.

At the same time, long term near-infrared (e.g. VVV) 
and mid-infrared (e.g. NEOWISE) 
time domain survey data has become available, leading to a plethora of outburst candidates.   
Only some of these candidates are amenable to the spectroscopic follow-up that would support an FU Ori classification, however.
And true confirmation would require multi-wavelength spectroscopy in order to detect the temperature and velocity gradients
expected from an accretion disk-dominated system.    In many of these cases of large or moderate-amplitude infrared brightenings, 
only a $K$-band spectrum, or perhaps even just a lightcurve, is available.

It thus becomes important to understand if there are photometric criteria that can be used to discriminate between 
FU Ori and other types of young star variable phenomena (as well as non-YSO contaminants).    
The mid-infrared expectations are particularly important to establish, 
as the threshold for declaring an outburst event between e.g. Spitzer and WISE observations separated by up to 5-7 years,
or in NEOWISE data that now spans 7 years, is only 1.5 to 2 mag \citep{fischer19,park21}.  
This is a mere factor of $\sim 5$ brightness increase, 
rather than the standard factor of $\sim$100 that is typically sought in optical searches.

For a disk outburst scenario, luminosity increases are a straightforward consequence 
of increases in disk accretion rate, perhaps combined with a change in the geometrical flow 
from being along magnetic fields, to equatorial.
For large enough outburst accretion rates, the entire ultraviolet, optical, near-infrared, and potentially mid-infrared
(depending on the radial extent of the outbursting region) wavelengths should be disk-dominated.   
Far-infrared and millimeter wavelengths are sensitive to reprocessed emission from the high-accretion zone.

The experiment we perform here is to contrast hypothetical low-state accretion disks, with a stereotypical high-state disk, 
and quantify expected outburst amplitudes as a function of wavelength.   

\section{Predicted Outburst Amplitudes}

Our toy model consists in the low state of:
\begin{itemize}
\item
A progenitor low-mass pre-main sequence star, with assumed values of  $T_{eff,*} =3800$ K and  $R_* = 1.5 ~R_\odot$ 
for its temperature and radius.  The SED is specified by a NextGen photosphere 
having luminosity $L_* = 4\pi R_*^2 \sigma T_{eff,*}^4$;
\item
A dust disk with inner radius corresponding to an assumed temperature of dust destruction, $T_{max,\ dust} = 1400$ K.  The SED
is modelled under the optically thick assumption, as a summation of blackbodies 
from the different radial annuli. The passive disk luminosity is $L_{dust}= 0.25 L_*$; 
\item 
A gaseous accretion disk with inner radius fixed at $2 ~R_\odot$.  Although the magnetospherically defined inner truncation radius 
would vary as a function of accretion rate, we keep this value constant for simplicity. 
The total accretion luminosity is given by $L_{gas,\ acc} = G M_* \dot M / R_*$ 
with a fraction up to 1/2 of this radiated by the disk, depending on the accretion flow geometry.
\end{itemize}

Our model in the high state consists of:
\begin{itemize}
\item
A pure accretion disk as presented in \cite{rodriguez2021}. The fiducial case here is a disk with maximum temperature 
$T_{gas,\ max} = 7000$ K and accretion rate of $\dot M = 10^{-5} ~M_\odot$/yr.
As above, 1/2 of the accretion luminosity is assumed to be radiated by the disk\footnote{The remainder 
is probably released in an extended ``boundary region" which differs from a radially thin classical boundary layer \citep{popham1996}.}. 
\end{itemize}
 
Contributions to the pre-outburst total source luminosity are $L_* = 0.42 ~L_\odot$, $L_{dust} = 0.10 ~L_\odot$, 
and $L_{gas,\ acc} = 0.046, 0.46, 4.6 ~L_\odot$ 
for corresponding accretion rates of  $\dot M = 10^{-8}, 10^{-7}, \textrm{and\ } 10^{-6} ~M_\odot$/yr, respectively.    
The post-outburst disk has $L_{gas,\ acc} = 40 ~L_\odot$, which is 100 times $L_*$. 

\begin{figure}
\centering
\includegraphics[scale=0.32]{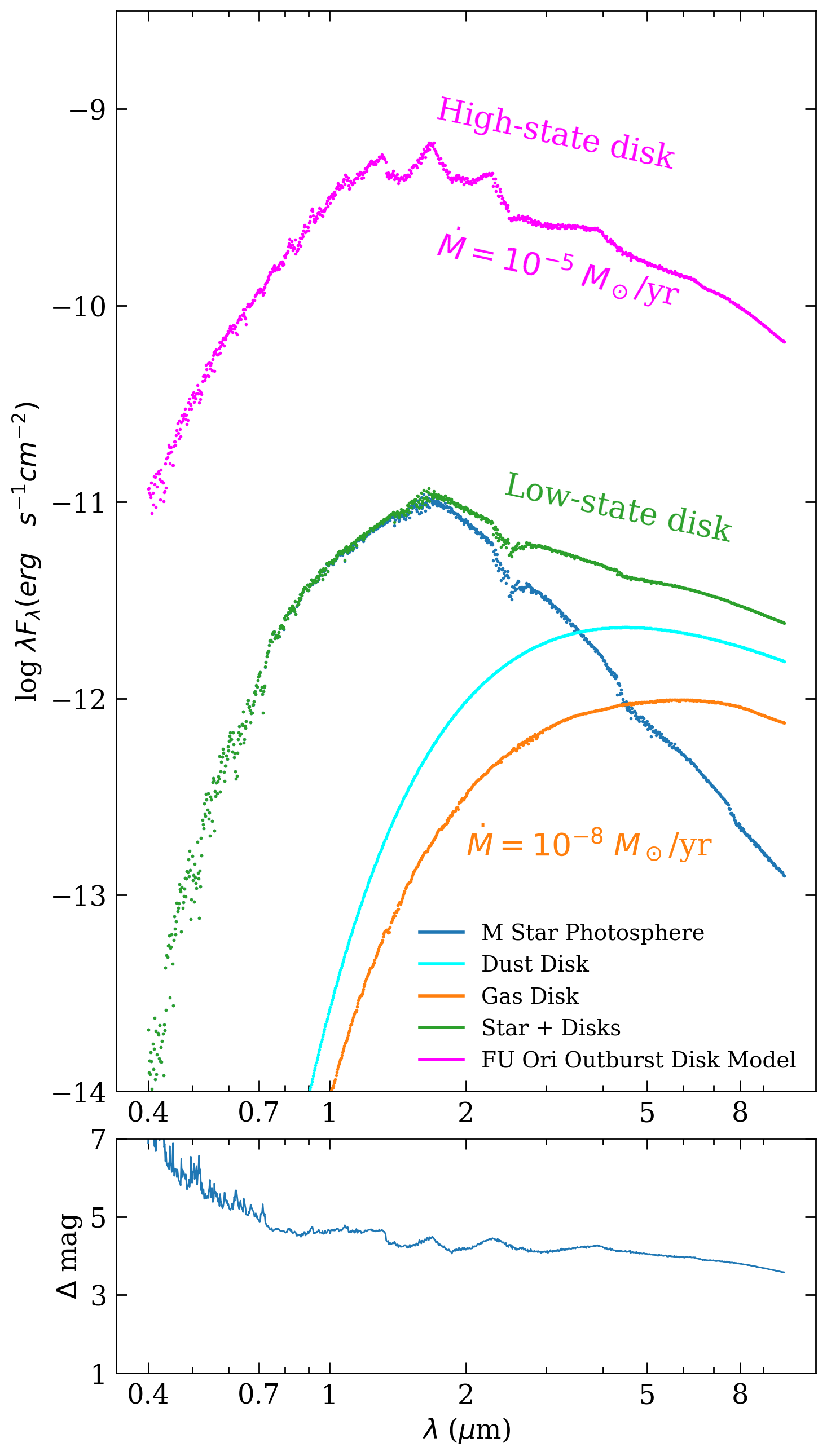}\includegraphics[scale=0.32]{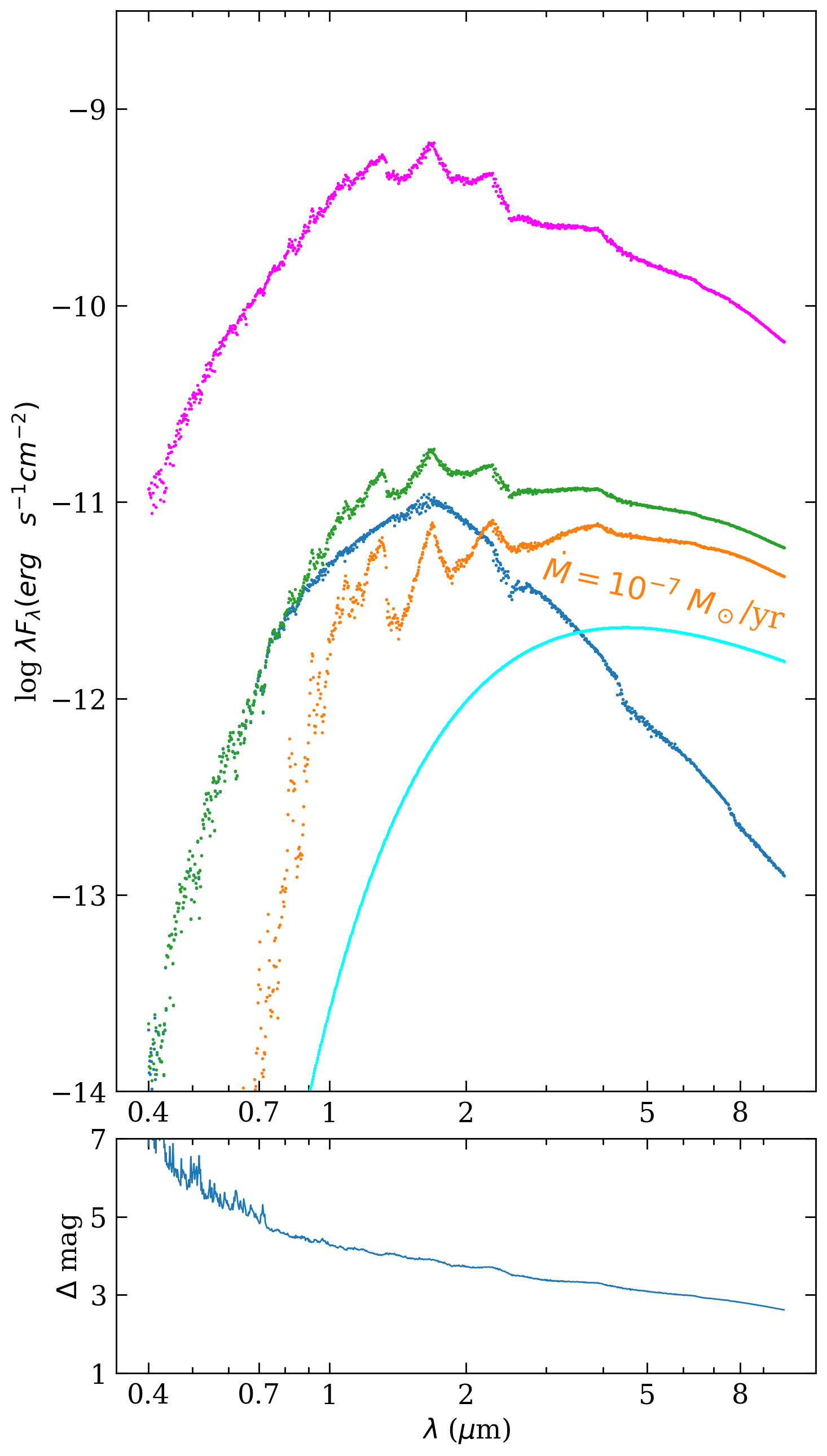}\includegraphics[scale=0.32]{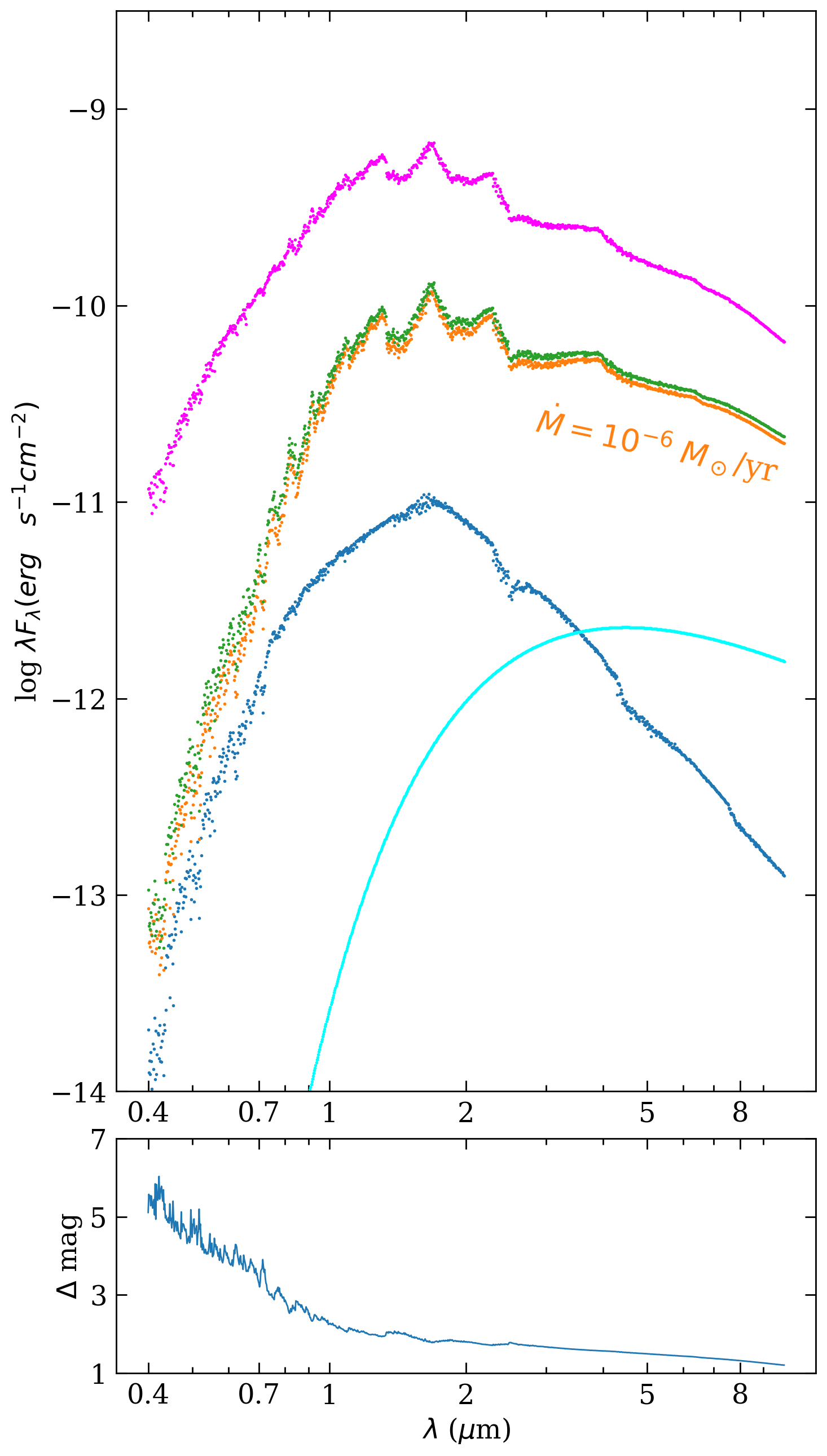}
\caption{{\bf Top panels:} Bright-state spectral energy distribution for an accretion disk-dominated FU Ori type system, 
accreting at $\dot M = 10^{-5} ~M_\odot$/yr (magenta), compared to composite spectral energy distributions (green) for
passive $+$ accretion disk systems having low-state accretion rates of $\dot M = 10^{-8},  10^{-7}, 10^{-6} ~M_\odot$/yr. 
In each panel, the same underlying stellar photosphere (navy blue), and passive dust reprocessing disk (cyan) are shown, 
along with the active gaseous accretion disk (orange); the assumed distance is $d=800$ pc.
{\bf Bottom panels:} Expected wavelength-dependent amplitude of an accretion outburst, calculated simply as the flux ratio of the magenta
and the green curves, converted to magnitudes.   As the low-state accretion rate increases, the outburst amplitudes decrease,
and the wavelength dependence of the brightening steepens. 
}
\label{ampwave}
\end{figure}

In Figure~\ref{ampwave} we show the pre-outburst and post-outburst disk model spectral energy distributions,
along with the predicted amplitudes of the outbursts.
As expected given the inner disk heating, the source brightening is larger amplitude towards the blue optical and ultraviolet.

The detailed trend with wavelength, 
from the red optical -- where most outbursts have been discovered -- to the mid-infrared -- where the most
comprehensive and uniform data set exists in NEOWISE, 
strongly depends on the low-state $\dot M$. 
For our fiducial star, the amplitudes are relatively flat with wavelength in the lowest $\dot M$ case,
typical of low-mass T Tauri stars in the Class II phase.
The amplitudes steepen towards the highest $\dot M$ case, which is more characteristic of the Class I phase.

The outburst amplitude behavior would also depend on 
the properties of the underlying central star, primarily its temperature,
and on the detailed temperature structure in the disk. 
We have adopted the classical $T(r)\propto r^{-3/4}$ distribution of a flat disk,
whereas a flatter profile such as that for a flared disk
would reduce the constrast, most significantly at the longer wavelengths $>$3 $\mu$m. 

\section{Summary and Implications}
We have provided a simple guide to expected outburst amplitudes 
as a function of wavelength for episodically occuring accretion-state transitions in YSOs,
from T Tauri type to FU Ori-like disk accretion.  
Consistent with traditionally quoted values from lightcurves, 
the models show 4-6 mag blue-optical outburst amplitudes 
and 1.5-4 mag mid-infrared amplitudes. 
Observation of such wavelength-dependent trends can help confirm FU Ori events 
in the absence of more secure diagnostics like high dispersion spectroscopy.


\begin{acknowledgements}
We thank Lee Hartmann and Will Fischer for giving this short contribution a once-over.
\end{acknowledgements}

\newpage
\bibliography{adstex}{}
\bibliographystyle{aasjournal}

\end{document}